\documentclass{iaus}
\usepackage{graphics}

  \checkfont{eurm10}
  \iffontfound
    \IfFileExists{upmath.sty}
      {\typeout{^^JFound AMS Euler Roman fonts on the system,
                   using the 'upmath' package.^^J}%
       \usepackage{upmath}}
      {\typeout{^^JFound AMS Euler Roman fonts on the system, but you
                   dont seem to have the}%
       \typeout{'upmath' package installed. iaus.cls can take advantage
                 of these fonts,^^Jif you use 'upmath' package.^^J}%
      }
  \else
  \fi

  \checkfont{msam10}
  \iffontfound
    \IfFileExists{amssymb.sty}
      {\typeout{^^JFound AMS Symbol fonts on the system, using the
                'amssymb' package.^^J}%
       \usepackage{amssymb}%

      }{}
  \fi

  \IfFileExists{amsbsy.sty}
    {\typeout{^^JFound the 'amsbsy' package on the system, using it.^^J}%
     \usepackage{amsbsy}}
    {\providecommand\boldsymbol[1]{\mbox{\boldmath $##1$}}}

\def\beq{\begin{equation}}
\def\eeq{\end{equation}}
\def\bey{\begin{eqnarray}}
\def\eey{\end{eqnarray}}
\def\pppm{\rm P^3M}
\def\mpc{\,h^{-1}{\rm {Mpc}}}
\def\mpci{\,h{\rm {Mpc}}^{-1}}

\def\kms{\,{\rm {km\, s^{-1}}}}
\def\msun{{M_\odot}}

\def\gs{\mathrel{\raise1.16pt\hbox{$>$}\kern-7.0pt
\lower3.06pt\hbox{{$\scriptstyle \sim$}}}}
\def\ls{\mathrel{\raise1.16pt\hbox{$<$}\kern-7.0pt
\lower3.06pt\hbox{{$\scriptstyle \sim$}}}}
\def\gtsima{$\; \buildrel > \over \sim \;$}
\def\ltsima{$\; \buildrel < \over \sim \;$}
\def\prosima{$\; \buildrel \propto \over \sim \;$}
\def\gsim{\lower.5ex\hbox{\gtsima}}
\def\lsim{\lower.5ex\hbox{\ltsima}}
\def\simgt{\lower.5ex\hbox{\gtsima}}
\def\simlt{\lower.5ex\hbox{\ltsima}}
\def\simpr{\lower.5ex\hbox{\prosima}}

\newsavebox{\astrutbox}
\sbox{\astrutbox}{\rule[-5pt]{0pt}{20pt}}

\title[Outskirts of Galaxy Clusters: intense life in the suburbs]
      {Velocity of galaxies with different luminosity}

\author[Y.P. Jing {\it et al.\/}]%
{Y.P. Jing$^1$ \and G. B\"orner$^2$}

\affiliation{$^1$Shanghai Astronomical Observatory, the Partner Group
of MPI f\"ur Astrophysik, \\Nandan Road 80, Shanghai 200030, China
email: ypjing@shao.ac.cn\\[\affilskip] 
$^2$ Max-Planck-Institut f\"ur Astrophysik,
Karl-Schwarzschild-Strasse 1, \\ 85748 Garching, Germany email: grb@mpa-garching.mpg.de}

\pubyear{2004}
\volume{195}
\pagerange{1--8}
\date{?? and in revised form ??}
\jname{Outskirts of Galaxy Clusters: intense life in the suburbs}
\editors{A. Diaferio, ed.}
\begin{document}

\maketitle

\begin{abstract}
We present the first determination of the pairwise velocity dispersion
of galaxies at different luminosity with the final release of the
Two-Degree Field Galaxy Redshift Survey (2dFGRS). Our result
surprisingly shows that the random velocities of the faint galaxies
are very high, around $ 700 \kms$, reaching similar values as the
brightest galaxies. At intermediate luminosities slightly brighter
than the characteristic luminosity $M_\star$, the velocities exhibit a
well defined steep minimum near $ 400 \kms$. The result challenges the
current halo model of galaxies of Yang et al. that was obtained by
matching the clustering and luminosity function of 2dFGRS, and can be
an important constraint in general on theories of galaxy formation,
e.g., the semi-analytical model. Combining the observed luminosity
dependence of clustering, our result implies that quite a fraction of
faint galaxies are in massive halos of galaxy clusters as the
brightest ones, but most of the $M_\star$ galaxies are in galactic halos.
\end{abstract}

\firstsection % if your document starts with a section,
\section{Introduction}

The clustering of galaxies in the Universe is characterized by their
spatial positions, and their peculiar velocities which lead to
deviations of their motion from the pure Hubble flow.  The big
redshift surveys assembled in recent years by the diligent work of
many astronomers give angular positions and redshifts for large
numbers of galaxies. A rough 3D map can be obtained by placing the
galaxies at distances along the line of sight derived via Hubble's law
from their redshifts. The peculiar velocity, however, also contributes
to the redshift, and this leads to a misplacement of the galaxy away
from its true location.  The local gravitational field is the cause of
the peculiar motion.  and thus the redshift distortion in the galaxy
maps can give information on the underlying matter distribution.

The amplitude of the distortions can be estimated from the pair
distribution of galaxies.  For pairs of galaxies at distances much larger than
their separation, one can use a plane-parallel approximation, and
write for the power spectrum in redshift space,
\begin{equation}
P^{S} ({k, \mu})=P(k)\frac{(1+\beta
  \mu^2)^2}{1+\frac{1}{2}(\sigma_v(k)k\mu)^2}\,.
\label{fitting} 
\end{equation}
Here $\mu$ is the cosine of the angle between the wave vector and the
line of sight. The linear redshift distortion parameter $\beta$ is
related to the linear growth factor $f(\Omega_M) \simeq \Omega^{0.6}$
($\Omega_M$ is the matter density), and the bias $b$ of the density
fluctuation spectrum via $\beta = \Omega_M^{0.6}/b$. P(k) is the power
spectrum in real space. The pairwise velocity dispersion (PVD)
$\sigma_v(k)$ describes the virial motion of galaxies in dense
systems, i.e. the Finger-of-God effect. As shown by Jing \& B\"orner
(2001), the redshift power spectrum of dark matter and its biased
tracers can be accurately described by Eq.(\ref{fitting}). Applying
this model to an observation of $P^S(k, \mu)$ may therefore yield a
determination of the three quantities $P(k)$, $\beta$, and
$\sigma_{v}(k)$, all of which are useful observables for testing
galaxy formation models

Here we use the final release of the 2dFGRS (
{\verb=http://www.mso.anu.edu.au/2dFGRS=}) to measure $P^{S} (k, \mu)$
and the PVD $\sigma_v(k)$ of galaxies. Since we are interested in the PVD
at small scale, we fix $\beta$ at a reasonable value of $ 0.45$
(Peacock et al. 2001).  The 2dF catalog has already been analyzed
statistically with respect to the PVD (Hawkins et al 2003), but here
we use a novel method to estimate the PVD. Furthermore, we take
advantage of the large number of galaxies in the 2dF catalog to bin
the galaxies in different luminosity intervals, and to make the first
study of the luminosity dependence of the PVD. This was not possible
up to now, and we shall see that remarkable tests of the galaxy
formation models become possible with the luminosity dependence of
the PVD.  The luminosity dependence of the clustering (the two-point
correlation function) of galaxies in the 2dFGRS has been
investigated (Norberg et al. 2002a), but not of the PVD.

\section{Statistical Analysis}

In order to study the luminosity dependence of the PVD, we divide the
galaxies into 11 subsamples according to their absolute
luminosity. The subsamples are successively brightened by 0.5
magnitude from the faintest sample $M_b=-17.0+5\log h$ to
$M_b=-22.0+5\log h$, with successive subsamples overlapping by 0.5
magnitude. Here $h$ is the Hubble constant in units of $100\kms{\rm
Mpc}^{-1}$.  For computing the absolute magnitude, we have used the
k-correction and luminosity evolution model of Norberg (2002b; ${\rm
k+e}$ model);, i.e., the absolute magnitude is in the rest frame $b_j$
band at $z=0$. We assume a cosmological model with the density
parameter $\Omega_0=0.3$ and the cosmological constant $\lambda_0=0.7$
throughout this paper.

We measure the redshift two-point correlation functions $\xi_z({\bf
s})$ following the method of Jing, Mo, \& B\"orner(1998).  The random
samples for the clustering analysis are generated in the same way as
described in Jing \& B\"orner (2004a). Each random sample for a
northern or southern luminosity subsample contains 100,000 random
points.  We convert the $\xi_z({\bf s})$ to $P^S(k,\mu)$ by the
Fourier transformation. More details about our statistical method can
be found in our journal paper (Jing \& B\"orner 2004b).

\section{Results}

\begin{figure}
\centerline{\resizebox{!}{12cm}{\includegraphics{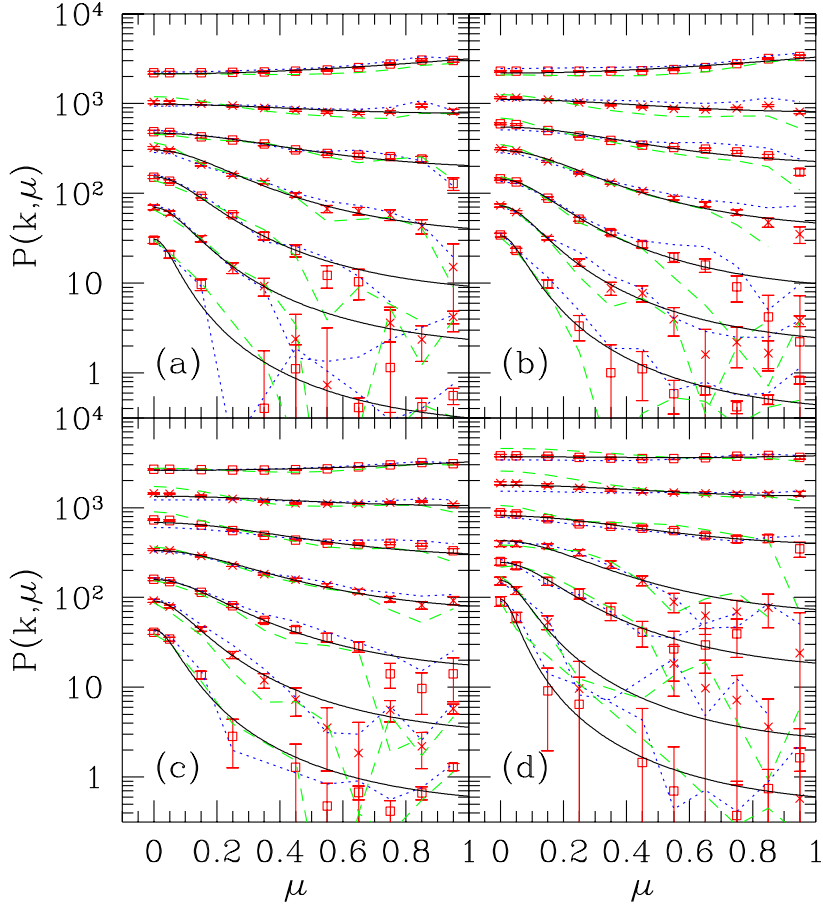}}}
%\center{\includegraphics[width=10cm]{fig5}}

\caption{The redshift space power spectrum $P(k,\mu)$ measured in
2dFGRS. The symbols are for the whole survey, the dotted lines for the
south subsample, and the dashed lines for the north subsample. The
errors are plotted only for the whole survey, estimated with
the bootstrap method. The smooth solid lines are the best fits of
Eq.(\ref{fitting}) to data of the whole sample. (a) for $-17.5<
M_b-5\log h <-18.5$; (b) for $-18.5< M_b-5\log h <-19.5$; (c) for
$-19.5< M_b-5\log h <-20.5$; (d) for $-20.5< M_b-5\log h <-21.5$.  The
$k$ values in each panel are from $0.16\mpci$ (top) to $2.5\mpci$
(bottom). }

\end{figure}

\begin{figure}
\centerline{\resizebox{!}{12cm}{\includegraphics{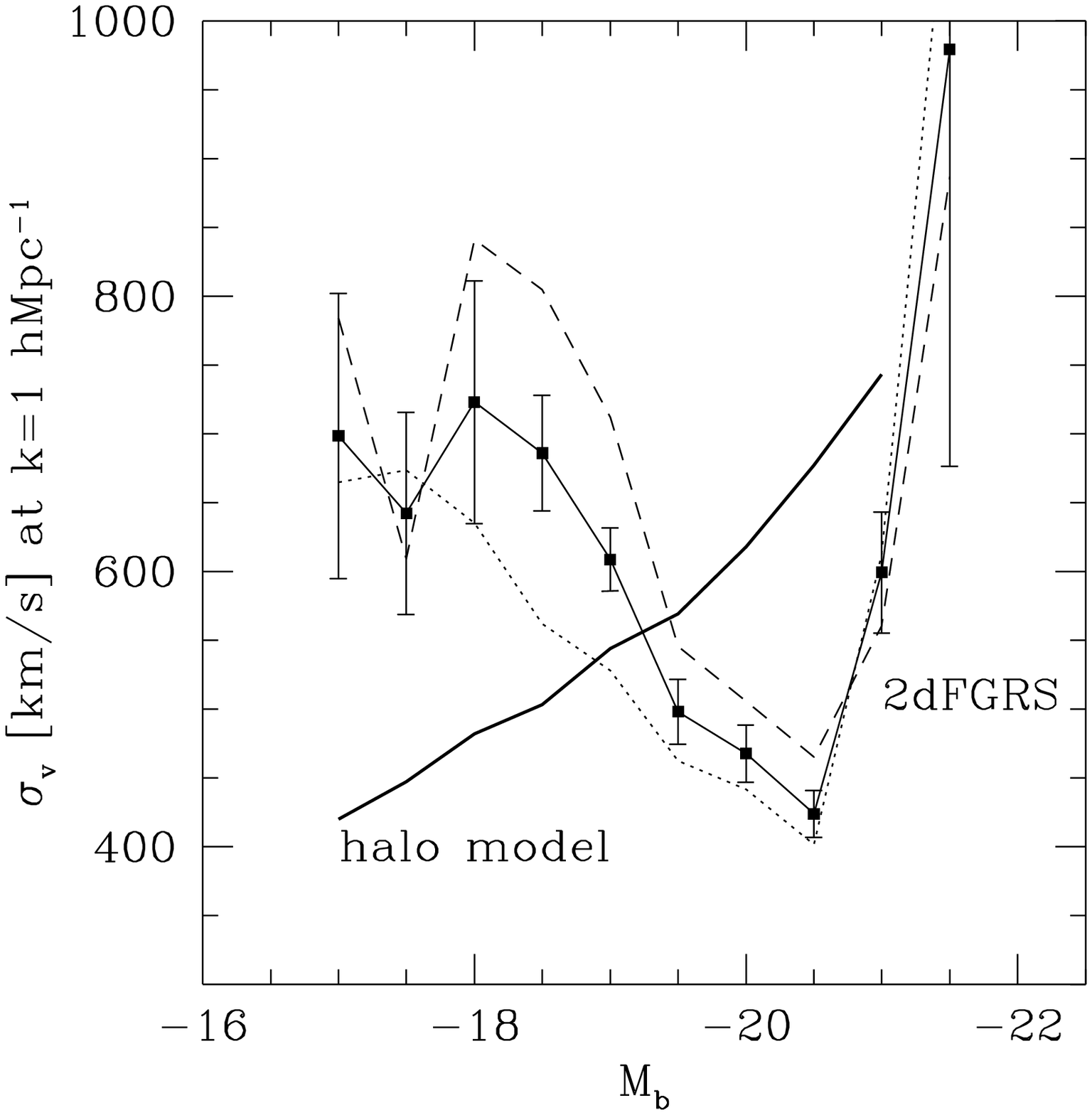}}}
\caption{The pairwise velocity dispersion measured at $k=1\mpci$ in
the 2dFGRS (symbols for the whole sample, dotted line for the south,
and dashed line for the north), compared with the prediction based on
the halo model (thick solid line). The error bars of the observed
results are given by the mock samples. }
\label{fig:sigmavk1}\end{figure}

In Fig.1 the basic measurement of the power spectrum in redshift space
$P(k, \mu)$ is shown. The four panels in this figure correspond to
four different luminosity intervals from faint to bright galaxies.
The values of k range from $0.16\mpci$ at the top to $2.5\mpci$ at the
bottom with an interval $\Delta \log_{10}k=0.2$. The south and north
samples agree quite well with the full survey indicating that cosmic
variance is not a major problem.  The power spectrum for the larger
k-values decreases quite strongly with $\mu$, more than a factor of
$10$ between $\mu = 0$ and $\mu = 1$.  The solid lines are the best
fits obtained by applying equation (\ref{fitting}) to the data of the
whole survey, showing that the model (\ref{fitting}) can accurately
describe the observed redshift power spectrum of galaxies.

The luminosity dependence of the PVD is shown most clearly in Fig.2,
where we have plotted $\sigma_{v}$ at $ k= 1 \mpci $ for the 11
overlapping samples. The surprising result is a strong dependence on
luminosity with the bright and the faint galaxies reaching high values
of $ \simeq 700 \kms$ or more, and a well defined minimum of $ \simeq
400 \kms$ for the galaxies of magnitude $M_b-5\log_{10} h =-20.5$. The
bright and the faint galaxies apparently have high random motions, as
expected for objects in massive halos or in clusters. The $M_\star$
like galaxies are rather moderate in their PVD, and probably reside in
galaxy size halos.

The thick solid line represents the prediction based on the up-to-date
halo model of Yang et al. (2003).  For the halo model, we adopt the
cosmological model that is a currently popular flat low-density model
with the density parameter $\Omega_0=0.3$ and the cosmological
constant $\lambda_0=0.7$ (LCDM). The shape parameter $\Gamma=\Omega_0
h$ and the amplitude $\sigma_8$ of the linear density power spectrum
are 0.2 and 0.9 respectively.  We use two sets of simulations, with
boxsizes $L=100\mpc$ and $L=300\mpc$, that were generated with our
vectorized-parallel $\pppm$ code (Jing \& Suto 2002) for this
model. Both simulations use $512^3$ particles, so the particle mass
$m_p$ is $6.2\times 10^8\msun$ and $1.7\times 10^{10}\msun$
respectively in these two cases.  We populate the halos with galaxies
in a similar way to that of Yang et al. (2004), but we used our own
code and adopted the model parameters of Model M1 in Yang et al. (2003).
The halo model, which matched nicely the luminosity function and the
luminosity dependence of clustering of galaxies in 2dFGRS (confirmed
by our analysis), clearly does not match our observation of the PVD.  The
failure of the halo model indicates that the prescription of how to
populate halos must be adapted better.

\section{Discussion}
The analysis of the velocity fields of the galaxies in the 2dFGRS has
led to a surprising discovery: The random velocities of the faint
galaxies are very high, around $ 700 \kms$, reaching similar values as
the bright galaxies. At intermediate luminosities the velocities
exhibit a well defined steep minimum near $ 400 \kms$.

It seems that the galaxies in different luminosity intervals appear as
different populations in their own right, defined by objective
statistics.  A look at figure 2 shows convincingly that this is
actually the case. For this figure we have sorted the galaxies in 11
luminosity bins, each one magnitude wide, from magnitude $-16.5$ to $
-22$ and plotted the value of $ \sigma_{v}$ at a wave number of $k
\simeq 1 \mpci$. Such a finely resolved binning of galaxies in samples
of different luminosities is possible for the 2dFGRS, because it is
big enough to contain sufficiently many galaxies in each luminosity
class.

The PVD is an indicator of the depth of the local gravitational
potential. Therefore we find the interesting result that the bright
and the faint galaxies move in the strongest gravitational field. They
are in clusters, while the galaxies around magnitude $-20$, the $M_*$
galaxies in the Schechter luminosity function, populate the field.

The bimodal nature of the correlation between the PVD and luminosity may
be used as a stringent test of galaxy formation models.  We have
investigated the halo occupation model (Yang et al. 2003) which has
been optimally fitted to reproduce the luminosity function, and the
two-point correlation function of the 2dFGRS. If we adapt this model
to the PVD value of the $M_*$ galaxies, we see that it cannot give the
high values found for the fainter galaxies. The PVD values of the
model actually run opposite to the data and the model assigns smaller
values to the fainter galaxies.  This must mean that the assignment of
galaxies to the dark matter halos must be done in a more intricate way
as up to now.  The number of faint galaxies in clusters must be
increased substantially to at least recover the high PVD found for
them.  Also, the low value of $ 400 \kms $ found for the galaxies with
magnitude $ -20.5$ must mean that these galaxies reside in dark matter
halos of galactic size, The halo population model must be fitted with
a much more complex scheme of assigning galaxies to halos, if the
results shown in figure 2 are to be reproduced,

Another way, widely used, to connect dark matter to galaxies is the
semianalytic modeling, where the dark matter distributions obtained
from N-body simulations are supplemented with some of the physical
processes important in galaxy formation using semianalytic techniques.
A test of the PVD vs luminosity for this type of models will be the
aim of a subsequent paper.

\begin{acknowledgments}
JYP would like to thank the Max-Planck Institute f\"ur Astrophysik for
its warm hospitality, IAU for its partial travel support, and the SOC
for adjusting the talk time. The work is supported in part by NKBRSF
(G19990754), by NSFC (No.10125314), and by the CAS-MPG exchange
program.
\end{acknowledgments}

\end{document}